\documentclass{article}

\usepackage[preprint, nonatbib]{neurips_2024}
\usepackage{amsmath, amssymb, amsthm}

\usepackage[numbers]{natbib}
\usepackage[utf8]{inputenc} 
\usepackage[T1]{fontenc}    
\usepackage{hyperref}       
\usepackage{url}            
\usepackage{booktabs}       
\usepackage{amsfonts}       
\usepackage{nicefrac}       
\usepackage{microtype}      
\usepackage{xcolor}         
\usepackage{amsmath}

\usepackage{graphicx}
\usepackage{subfigure}
\usepackage{ulem}

\theoremstyle{plain}
\newtheorem{theorem}{Theorem}[section]
\newtheorem{proposition}[theorem]{Proposition}

\theoremstyle{definition}
\newtheorem{definition}[theorem]{Definition}

\theoremstyle{remark}
\newtheorem{remark}[theorem]{Remark}

\title{Quantum State Generation with Structure-Preserving Diffusion Model}

\author{%
   Yuchen Zhu \\ 
  Georgia Tech\\
  \And
  Tianrong Chen \\
  Georgia Tech \\
  \And
  Evangelos A. Theodorou \\
  Georgia Tech \\
  \And
  Xie Chen \\
  Caltech \\
  \And
  Molei Tao \\
  Georgia Tech \\
}

\begin{document}

\maketitle

\begin{abstract}
    This article considers the generative modeling of the (mixed) states of quantum systems, and an approach based on denoising diffusion model is proposed. The key contribution is an algorithmic innovation that respects the physical nature of quantum states. More precisely, the commonly used density matrix representation of mixed-state has to be complex-valued Hermitian, positive semi-definite, and trace one. Generic diffusion models, or other generative methods, may not be able to generate data that strictly satisfy these structural constraints, even if all training data do. To develop a machine learning algorithm that has physics hard-wired in, we leverage mirror diffusion and borrow the physical notion of von Neumann entropy to design a new map, for enabling strict structure-preserving generation. Both unconditional generation and conditional generation via classifier-free guidance are experimentally demonstrated efficacious, the latter enabling the design of new quantum states when generated on unseen labels.
    \end{abstract}
    \vskip -0.25in

    \section{Introduction}
    
    Generative modeling, a specialized subfield of machine learning, is dedicated to creating models capable of generating new data points that closely match the statistical distribution of a given dataset. These models adeptly capture and replicate the patterns, structures, and characteristics of the input data, allowing them to yield new instances that seem to originate from the same distribution. Among them, diffusion model \citep{sohl2015deep,ho2020denoising,song2020score} has risen to prominence for its remarkable ability to produce high-quality data, especially notable in image generation \citep{rombach2022high,ho2022cascaded}. The fundamental mechanism of diffusion model starts with adding noise to the data and progressively altering it into a tractable Gaussian prior with analytically available transition kernels. The model then effectively reverses this noise addition, reconstructing the data from its noised state through the application of learned score functions \citep{anderson1982reverse,haussmann1986time}. This process not only results in easily obtainable conditional score functions but also simplifies regression objectives, significantly boosting their capability in managing complex, high-dimensional data distributions. 
    
    Diffusion model has showcased impressive capabilities in generating various types of dataset across different application area, such as video synthesis \citep{ge2023preserve,blattmann2023align}, language modeling \citep{austin2021structured,lou2023discrete} and point cloud generation \citep{zeng2022lion,luo2021diffusion}. With the recent rise of AI4Science, the potential of diffusion model in tackling scientific problems also started being explored, including applications in biology \citep{jumper2021highly,watson2022broadly}, chemistry \citep{duan2023accurate,hoogeboom2022equivariant}, and climate science \citep{mardani2024residual}. Other important notions relevant to this article include constrained generations \citep[e.g.,][]{liu2023mirror,
    fishman2023metropolis,
    fishman2023diffusion,
    lou2023reflected} and conditional generations \citep[e.g.,][]{dhariwal2021diffusion,ho2022classifier}.

    \begin{figure*}[t!]
            \centering
            \label{fig:highlight}
            \begin{subfigure}
                \centering
                \includegraphics[width=\textwidth]{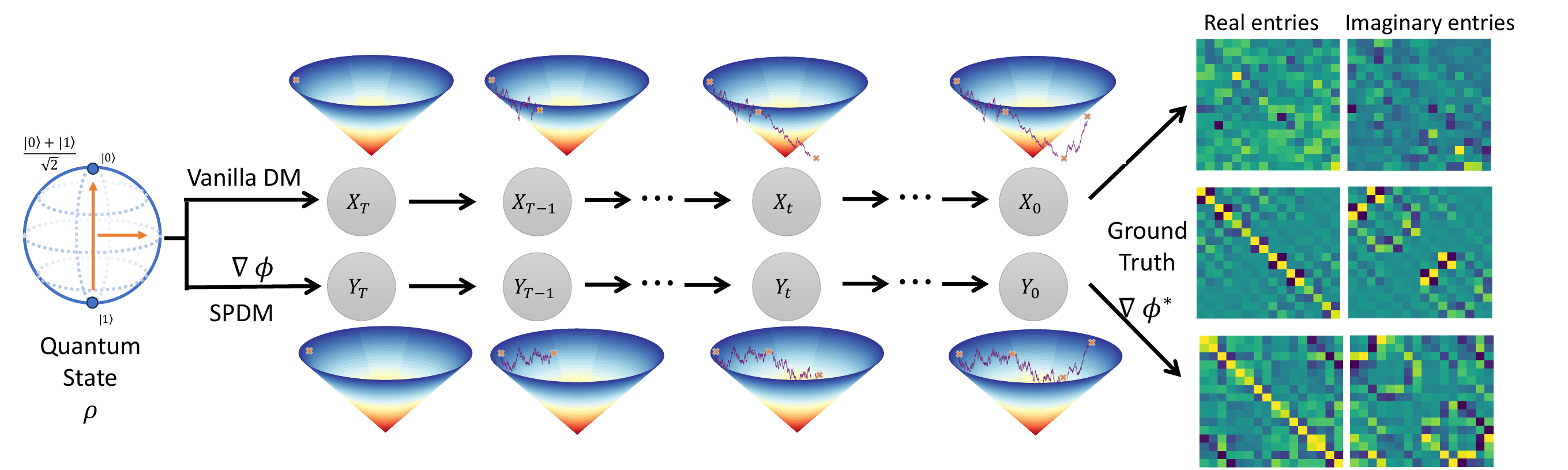}
            \end{subfigure}%
        \caption{Structure-Preserving Diffusion Model (SPDM) is the first diffusion-based method for quantum state generation. It hard codes physical knowledge into the generative models to strictly satisfy the structural conditions of quantum states (lower path; dynamics inside the constraint cone). In contrast, vanilla Diffusion model (DM) falls short of generating reliable samples due to its failure to detect the sophisticated structures of quantum states in high dimensions (upper path; dynamics ignores the constraint cone, leading to unsatisfactory generation quality).
        }%
        \end{figure*}
    What is under-studied but important is the (classical) generative modeling of quantum data. This is not the same as quantum computing or quantum machine learning, because the goal is not to design quantum algorithms, but to develop algorithms that run on classical computers for handling quantum data. 

    Similar to the already extremely successful application of the generation (and thus design) of molecular configurations \citep[e.g.,][]{watson2023novo, duan2023accurate}, it would be useful to generate, for example, the states of quantum systems, before quantum computers become accessible to general researchers.
    
    To be able to represent quantum states, which in general can be mixed states, one needs to be mindful of their structural constraints that originated from physics. For example, it is standard in quantum information and quantum computing to represent mixed states as complex-valued density matrices \citep{nielsen2010quantum}, but such matrices have to be Hermitian, positive semidefinite, and trace one. Data-based generation of quantum density matrices therefore amounts to generative modeling in non-Euclidean spaces. Without strictly satisfying the structural constraints, the generated matrices will no longer correspond to quantum states \citep[e.g.,][]{huang2020predicting}. Unfortunately, generic generative modeling approaches will almost surely violate the constraints due to indispensable errors originated from finite data and compute. This article proposes the first machine learning methodology that generates density matrices similar to training data and strictly satisfying their structural constraints, based on hard-wiring these constraints into diffusion models via technique from convex optimization.
    
    Imperative to mention is, machine learning has already been demonstrated promising for quantum problems, and it is rapidly becoming popular in the domain. There is even an interesting research frontier on quantum algorithms for generative modeling \citep[e.g.,][]{dallaire2018quantum,khoshaman2018quantum, chakrabarti2019quantum, stein2021qugan, zhang2024generative, chen2024quantum}; however, this article still focuses on classical algorithm as quantum computer is not yet widely available, and existing results did not generate \textbf{mixed} states larger than 3-qubits. Meanwhile, classical algorithms have already been considered for many important quantum problems, such as the classification and regression of quantum states \citep[e.g.,][]{huang2020predicting, huang2022provably} and quantum state tomography \citep[e.g.,][]{torlai2018neural, torlai2018latent, carrasquilla2019reconstructing, vicentini2022positive, jayakumar2023learning}. The latter, roughly speaking, corresponds to the process of estimating one latent quantum density matrix through repeated, but necessarily noisy measurements of many copies of the same state. It is very important to clarify  the similarity and difference between this work and quantum state tomography:
    
    In fact, some may feel that quantum state tomography is already generative modeling, in a broad sense, because once the latent deterministic density matrix is recovered from noisy measurements, one can produce additional, synthetic classical measurements, which are random and ideally identically distributed with the training data. However, it is actually very different from the generative modeling of quantum states considered here, which might be in a narrow sense, however being a common setup in (classical) machine learning. More precisely, what this article considers is a collection of many \textbf{different} quantum mixed states, instead of classical observations of a \textbf{single} quantum state, and the goal is to generate more mixed states that are similar to the given ones. The problem of using machine learning to infer a single quantum state from classical observational data, i.e. state tomography, has already been considered by a series of seminal work mentioned above; there the probability distribution of data is uniquely determined by a latent quantum state, and thus one essentially needs to learn a $D$-dimensional \textbf{vector} (if the latent state is an $N$-qubit mixed state represented by a $2^N\times 2^N$ density matrix, then $D=2^{2N}$). The new task of state generative modeling in this article, on the other hand, assumes mixed states are already given and they follow some unknown latent probability distribution, which can be represented by a probability density function of $D$ variables, and the goal is in some sense to learn this \textbf{function}, which is exponentially more complex than a $D$-dimensional vector. One extra difference between this work and existing milestones on generative modeling for state tomography is, from a machine learning methodology point of view, this work is based on diffusion for generative modeling, while work cited above used, e.g., EBM, RBM, RNN instead.
    
    What is the point of generating more quantum states?
    
    One perspective is to actually create innovative states. Unconditional generation will simply produce states similar to those in the training data, but modern (diffusion) generative modeling technique also unleashes the ability of conditional generation. More precisely, given training data that carry different labels (e.g., traditionally labels could be `astronaut', `horse', etc., but in our case they can be `insulating', `magnetic', `topological', etc.), we can not only generate under a label that already has training data (e.g., `astronaut', or `magnetic'), but also generate under a label that has no training data (e.g., the famous `horse-riding astronaut' \citep{HorseRidingAstronaut}, or, this time, a motivating perspective of creating `topological magnetic' states). This article humbly starts a first step toward this goal of scientific innovation, essentially via the interpolation/extrapolation of quantum data. The inter-/extra-polation, however, is in the sense of probability distributions but not sample points, because when one tries to design new states, it may not make sense to take, e.g., a specific `topological' state and a specific `magnetic' state and force a new `topological magnetic' state based on them two alone. Instead, our conditional generative modeling approach hybrids multiple probability distributions and innovate just like prevailing generative AI tools.
    
    Another application is, generative modeling can also estimate the probability distribution of states in the data, hence providing further quantitative insights for probing the physical process behind it. Some consideration of this aspect was already included in the state tomography literature \citep[e.g.,][]{carrasquilla2019reconstructing}, but not for an ensemble of states. Moreover, diffusion model excels in this regard. As a more recent generative approach, it is posited to be particularly effective for estimating density near the data manifold, even when data is high-dimensional but sample size is small, as long as there is a low-dimensional structure in the data. This is oftentimes the case for image and video generations, and their empirical successes were the root of this belief. Quantum density matrices data could be similar; at least, they are high-dimensional since dimension grows exponentially with the number of qubits, but scarce due to the cost of experiments. This is a case where diffusion generative model could be advantageous over traditional approaches. 
    
    In addition, generative model can serve as a sampler for crafting synthetic data that are experimentally expensive to harvest. Synthetic data that are statistically similar to training data are useful for discovering system properties through helping estimate observable values. The seminal work \citep{carrasquilla2019reconstructing} showcases one such application for state-tomography type generative modeling. The idea is, generative modeling could help obtain more virtual `measurements' from physical ones, hence improving the accuracy of the estimated latent state. Analogously, generative modeling of a distribution of mixed states can help the investigation of stochastic quantum systems, such as Random Transverse-field Ising Models (RTFIM) \citep{fisher1992random, fisher1995critical, kovacs2022quantum} and measurement-altered quantum critical systems \citep{murciano2023measurement}. The generated states can empower more accurate estimation of random parameters in the system, such as coupling and transverse field strengths.

    \vspace{-0.2cm}
    \section{Background}
    \vspace{-0.1cm}
    \subsection{Unconditional Generation via Diffusion Model}
    \vspace{-0.1cm}
    Denoising diffusion generative models in Euclidean space admit many descriptions, with focuses on different perspectives, and here we adopt the stochastic differential equations (SDE) description \citep{song2020score}. Given samples of $\mathbb{R}^{d}$-valued random variable $x_0$ that follows the data distribution $p_0$, denoising diffusion adopts a forward noising process followed by a backward denoising generation process to generate more samples of $p_0$. 
    
    The forward process transports the (unknown latent) data distribution to a known, easy-to-sample distribution by evolving the initial condition via an SDE,
    \begin{align}
    \label{eq:forward-sde}
        \mathrm{d} x_t = f(x_t, t) \mathrm{d}t + g(t) \mathrm{d}w_t .
    \end{align}
    The evolution of the distribution's density $p_t$, given by $x_t \sim p_t(\cdot)$, can be characterized by a partial differential equation (PDE) known as
    the Fokker-Planck Equation (FPE),
    \begin{align*}
        \frac{\partial}{\partial t}p_t(x) = -\operatorname{div}(p_t(x) f(x,t)) + \frac{g(t)^2}{2} \Delta p_t(x).
    \end{align*}
    Upon certain choices of the drift $f$ and diffusion coefficient $g$, the solution to the FPE will approach some limiting distribution. For example,
    $f(x, t) = - x, g(t) = \sqrt{2}$ corresponds to the well-known Variance-Preserving (VP) scheme, also known as Ornstein–Uhlenbeck process. In this case, $p_\infty$ will be a standard Gaussian $\mathcal{N}(0, I)$. 
    
    The backward process then utilizes the time-reversal of the SDE \eqref{eq:forward-sde} \citep{anderson1982reverse}. More precisely, if one considers
    \begin{align}
    \label{eq:backward-sde}
    \mathrm{d}y_\tau &= (-f(y_\tau, T-\tau) + g(T-\tau)^2 s(y_\tau, T-\tau) ) \mathrm{d} \tau + g(T-\tau) \mathrm{d} w_\tau,
    \end{align}
    with $y_0 \sim p_T$, where $s$, known as the score function, is given by $s(x,t)=\nabla \log p_t (x)$,
    then we have $y_\tau \sim p_{T-\tau}$, i.e. $y_\tau = x_{T-\tau}$ in distribution. In particular, the $T$-time evolution of \eqref{eq:backward-sde}, $y_T$, will follow the data distribution $p_0$.

    In practice, one considers evolving the forward dynamics for finite but large time $T$, so that $p_{T} \approx \mathcal{N}(0, \sigma_{T}^2 I)$, and then initialize the backward dynamics using $y_0 \sim \mathcal{N}(0, \sigma_{T}^2 I)$ and simulate it numerically till $\tau=T$ to obtain approximate samples of the data distribution. Critically, the score function $s$ needs to be estimated in the forward process.

    To do so, the score $s(x,t)=\nabla \log p_{t}(x)$ is often approximated using a neural network $s_{\theta}(x, t)$. For linear forward SDE, it is typically trained by minimizing an objective based on denoising score matching \citep{vincent2011connection}, namely
    \begin{align*}
        \mathbb{E}_{t} \mathbb{E}_{x_0 \sim p_0} \mathbb{E}_{x_t \sim p_t(\cdot|x_0)} \| s_{\theta}(x_t, t) -\nabla \log p_t(x_t | x_0)\|^2 
    \end{align*}
    where $\nabla \log p_t(x_t | x_0)$ is the conditional score derived from 
    \eqref{eq:forward-sde} with a given initial condition.     
    
    \subsection{Diffusion Guidance for Conditional Generation}
    For the task of conditional generation, we have an additional input $y$ which is often a class label (e.g., a text sequence), and the goal is to sample from the conditional distribution $p(x|y)$ given training data $\{x_i,y_i\}_i$. For diffusion model, this means that we need the conditional score function $\nabla \log p(\cdot | y)$. After applying Bayes' rule, it's clear that we can decompose the conditional score into two parts,
    \begin{align*}
        \nabla \log p(x | y) = \nabla \log p(x) + \nabla \log p(y|x),
    \end{align*}
    where $\nabla \log p(x)$ is the usual score function of the data distribution, and $\nabla \log p(y|x)$ is the gradient of the conditional probability of the addition input being $y$. $\nabla \log p(x)$ can be learned by training a diffusion model on unconditional data, and we need an estimator of $\nabla \log p(y|x)$ to generate samples from the conditional distribution $p(\cdot|y)$. Noticed that $p(y|x)$ can be approximated by training a discriminative model such as a classifier based on the data-label pair $(x,y)$. In practice, practitioners often use pre-tained classifier models to estimate $\nabla \log p(y|x)$.
    
    Dhariwal and Nichol \citep{dhariwal2021diffusion} exploited this fact and proposed the technique of classifier-based diffusion guidance, originally for boosting the sample quality generated by diffusion models. Instead of $p(x|y)$, they proposed to sample from a condition-enhanced distribution $p_{\gamma}(x | y) \propto p(x) p(y|x)^{\gamma}$.
    The score of this condition-enhanced distribution can be computed by 
    \begin{align*}
        \nabla \log p_{\gamma}(x|y) = \nabla \log p(x)  + \gamma \nabla \log p(y|x).
    \end{align*}
    $\gamma$ is called the strength of the guidance, which amplifies the influence of the conditioning when setting to a scale larger than $1$. When $\gamma > 1$, the distribution is sharpened and focused onto its mode that corresponds to the condition $y$ \citep{dieleman2022guidance}. By tuning $\gamma$, classifier guidance allows us to capture the influence of the condition signal. 
    
    However, classifier-based diffusion guidance can become impractical due to the expensive cost of training a separate classifier model. Such a classifier often requires to be noise-robust to handle the noise-corrupted input $x$ during the sampling process. \citep{ho2022classifier} proposed classifier-free diffusion guidance to sample from the condition-enhanced distribution $p_{\gamma}(x|y)$  without the need for an extra discriminative model. Their insight came from another writing of Bayes' rule: $\nabla \log p_{\gamma}(x|y)$ can also be decomposed as,
    \begin{align*}
        \nabla \log p_{\gamma}(x | y) = (1 - \gamma) \nabla \log p(x) + \gamma \nabla \log p(x|y).
    \end{align*}
    Therefore, classifier-free diffusion guidance motivates the training of a model that functions both as the conditional score $\nabla \log p(x|y)$ and the unconditional score $\nabla \log p(x)$, depending on whether an additional condition input $y$ is given. We will use classifier-free guidance due to improved performance and computational cost.

    \subsection{Density Matrix Representation of Quantum State}
    Since we will consider the unconditional and conditional generations of quantum states, let's review some of their basics. A quantum state describes the physical status of a quantum system. For generality and with some motivations being quantum computing and quantum information in mind, we consider mixed states, which can be represented by density matrices (in theory, they are operators, but once we fix the bases of the Hilbert space, they become finite dimensional objects).
    
    Precisely, a (quantum) density matrix is a matrix with each element being a \textbf{complex} number, which satisfies the following structural constraints \citep[e.g.,][]{nielsen2010quantum}:
    
    It is Hermitian, with trace 1, and positive semi-definite.
    
    \textbf{Notations:} Here we denote the set of complex Hermitian matrices of dimension $n$ as $\mathbb{C}_n$, 
    \begin{align}
        \mathbb{C}_n = \{X \in \mathbb{C}^{n \times n}, X = X^{\dagger}\},
    \end{align}
    where $X^{\dagger}$ is the conjugate transpose of $X$, defined through $(X^{\dagger})_{ij} = \overline{X_{ji}}$.
    We also denoted the set of positive definite complex Hermitian matrices of dimension $n$ as $\mathbb{C}_{n}^{+}$
    \begin{align}
        \mathbb{C}^{+}_{n} = \{X \in \mathbb{C}_n, X \succ 0\}.
    \end{align}

    \begin{definition}
        A $n-$qubit quantum state $\rho$ with full rank is an element of $\mathbb{C}^{+}_{2^{n}}$ with trace 1.
    \end{definition}
    
    Because of the structural constraints, the generative modeling of quantum states is nontrivial but an interesting machine-learning problem. Ideally speaking, if there were an infinite amount of data and infinite computational resources (so that generative modeling can be exact), the training data already implicitly carry information about the structural constraints, and then the generated data will also respect the constraints, thus yielding true quantum states. Nevertheless, in practice the generation is plagued with various sources of error (e.g., statistical error of finite samples, (score) function approximation error, finite-time approximation of $T\to\infty$, numerical integration of the backward process, etc.) and generic generative modeling approaches that are not built to respect those constraints will likely generate samples that violate them, and the samples will no longer possess physical meaning and are hence useless. Therefore, in the following, we hardwire these important physical knowledge into our generative model, so that the constraints will be \textbf{exactly} satisfied.
    \vskip -0.4in
    
    \section{Method}
    Our task is to generate new samples of quantum mixed states, from a distribution characterized by the training data. Out of the three structure constraints of density matrix representation of quantum states, Hermitianity is easy to handle as one can just work with the upper triangular part of the matrix. Trace one condition will only be minimally violated due to generation error, and we simply normalize the generated result by dividing its trace to strictly enforce this constraint. The most nontrivial constraint is the positive semi-definiteness. Since this is a convex constraint, we can leverage the recent approach of Mirror Diffusion Model (MDM) \citep{liu2023mirror}, which is a class of diffusion models that generate data on convex-constrained sets, to enable exact generation on $\mathbb{C}^{+}_{n}$. To achieve this goal, we work out a complex and multivariate extension of the results in MDM.

    \subsection{Mirror Map with Real Numbers}
    More precisely, the approach of MDM relies on a tool from convex optimization known as mirror map. It can create a nonlinear bijection that maps the data from the constrained primal space to an unconstrained dual space, which is Euclidean. This bijection serves as a coordinate transformation, i.e. a pair of exact encoder and decoder. A mirror map $\nabla \phi$ is defined as follows:
    \begin{definition}
    Given a convex constrained set, $\mathcal{M} \subseteq \mathbb{R}^{d}$, $\nabla \phi$ is called mirror map if $\phi: \mathcal{M} \to \mathbb{R}$ is a twice differentiable function that is strictly convex and satisfying:
    \begin{align*}
        \nabla \phi(\mathcal{M}) = \mathbb{R}^{d} \quad \text{and} \quad \underset{x \to \partial \mathcal{M}}{\lim} \| \nabla \phi \| \to \infty.
    \end{align*}
    \end{definition}
    \vskip -0.2in
    $\nabla \phi: \mathcal{M} \to \mathbb{R}^{d}$ is an invertible map from the constrained primal space $\mathcal{M}$ to an unconstrained dual space $\nabla \phi(\mathcal{M})$, which is also called the mirror space.
    
    Its inverse enjoys a pleasant property from convex analysis \citep{rockafellar2015convex}: by defining the dual of $\phi$ as
    $
        \phi^{*}(y) = \sup_{x \in \mathcal{M}} \langle x, y \rangle - \phi(x),
    $
    the inverse (as mapping) of $\nabla \phi$ is the same as the gradient of $\phi^{*}$, i.e.
    \begin{align*}
        \nabla \phi^{*} (\nabla \phi(x)) = x, \ \forall x\in\mathcal{M}, \quad \nabla \phi (\nabla \phi^{*} (y)) = y,
        \ \forall y\in\mathbb{R}^d.
    \end{align*}
    \vskip -0.1in
    Therefore, $(\nabla \phi, \nabla \phi^{*})$ acts as a pair of encoder-decoder that transform each datum between the primal constrained space $\mathcal{M}$ and the unconstrained mirror space $\nabla \phi(\mathcal{M})$.

    \subsection{Mirror Map for Complex-Valued Density Matrices}
    A complication is that our data consists of complex-valued matrices, different from real vectors and matrices typically considered in convex optimization. Therefore, what remains to be described is the construction of a strictly convex potential $\phi$ for the class of positive definite Hermitian matrices. It turns out that the negative von Neumann entropy satisfies the need.

    The von Neumann entropy \citep[e.g.,][]{bengtsson2017geometry} is an extension of the concept of Shannon entropy to quantum states. 
    For $X \in \mathbb{C}_{n}^{+}$, the negative von Neumann entropy is defined as $\phi(X) = \operatorname{Tr}(X \log X)$.
    The mirror map and its inverse are computed analytically as
    \begin{align}
        \nabla \phi(X) = I + \log X, \quad \nabla \phi^{*}(Y) = \exp( Y - I),
        \label{eq:ourMirrorMaps}
    \end{align}
    where log and exp are matrix logarithm and matrix exponential. Specifically, we define the matrix logarithm $\log X$ to be
    $\log X = \sum_{i} (\log \lambda_i) q_{i} q_{i}^{\dagger}$ if $X$ admits spectral decomposition $X = \sum_{i} \lambda_i q_i q_i^{\dagger}$, where $\lambda_i > 0$ are positive real numbers, $\{q_i \}$ are a set of complex-valued vectors that are also orthonormal basis of $\mathbb{C}^{n}. $ 
    
    \begin{proposition}\label{prop:mirror_fn}
        $(\nabla \phi, \nabla \phi^*)$ given by \eqref{eq:ourMirrorMaps} is a pair of bijective mappings between the set of positive definite complex Hermitian matrices $\mathbb{C}_{n}^{+}$ and the set of complex Hermitian matrices $\mathbb{C}_n$. Moreover, $\nabla \phi^{*}$ is the inverse of $\nabla \phi$.
    \end{proposition}
    \vskip -0.1in
    \textit{Proof.} Please see Appendix.\ref{Appendix:prop:mirror_fn_proof}.
    
    \subsection{Generative Modeling in Primal Space Based on Diffusion in Dual Space}
    The gradient map of negative von Neumann entropy, $\nabla\phi$, allows us to transform the problem of learning and sampling the data distribution $p_{data}(X)$ over $\mathbb{C}_{n}^{+}$, to that for a new distribution $\tilde p_{data} = ([\nabla \phi_{\#}]p_{data})$ over the mirror space $\mathbb{C}_n$, where $\#$ is the push-forward operation. Moreover, although $\mathbb{C}_n \subseteq \mathbb{C}^{n \times n}$ is still a convex-constrained set due to each element being a Hermitian matrix, $\mathbb{C}_{n}$ enjoys a much simpler structure due to the fact that $\mathbb{C}_n \cong \mathbb{R}^{n^{2}}$, i.e. it is isomorphic to an unconstrained real Euclidean space of dimension $n^{2}$. This is because each complex Hermitian matrix can be represented using only its upper or lower triangular entries. Since complex Hermitian has real diagonal values, the diagonal contributes $n$ dimension. There are $n(n-1)/2$ strict upper triangular entries, each having a complex value. Each complex entry can be represented with a point in $\mathbb{R}^{2}$. Therefore, the total dimension is $n^2$ and linearity guarantees the isomorphism.
    
    Since diffusion models leverage a neural network parameterization of the score function, and most neural networks have not been designed for complex numbers, we utilize the isomorphism mentioned above and further transform $\nabla \phi(X)$ into a vector in $\mathbb{R}^{n^2}$. This enables us to represent the score function based on simple Euclidean space again.

    \begin{figure*}[!h]
        \centering
        \label{fig:uncond}
        \begin{subfigure}
            \centering
            \includegraphics[height=1.3in]{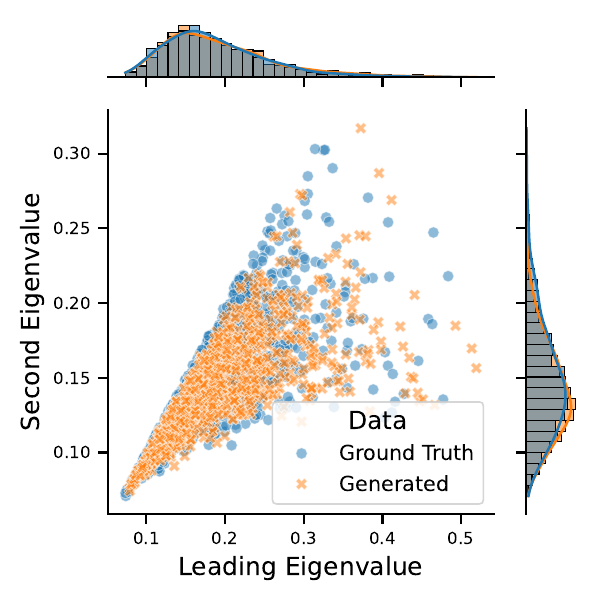}
        \end{subfigure}%
        ~ 
        \begin{subfigure}
            \centering
            \includegraphics[height=1.3in]{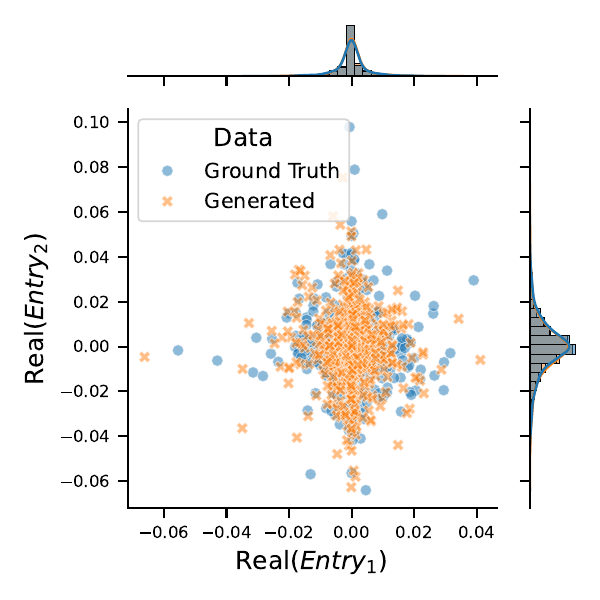}
        \end{subfigure}
        \begin{subfigure}
        \centering
        \includegraphics[height=1.3in]{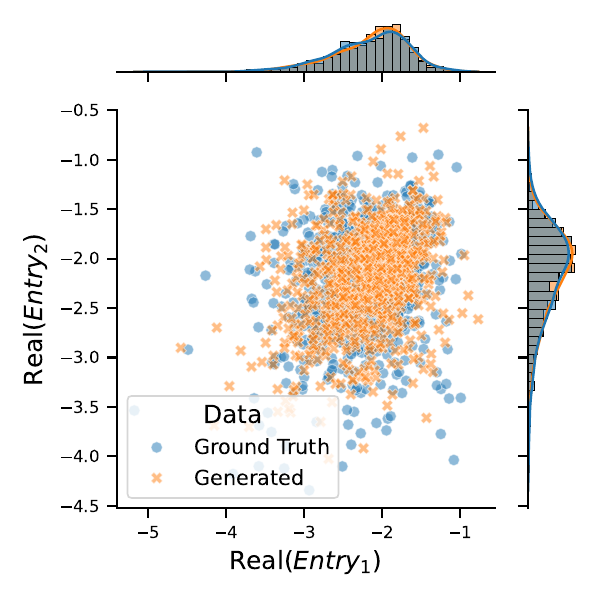}
        \end{subfigure}
        \begin{subfigure}
        \centering
        \includegraphics[height=1.2in]{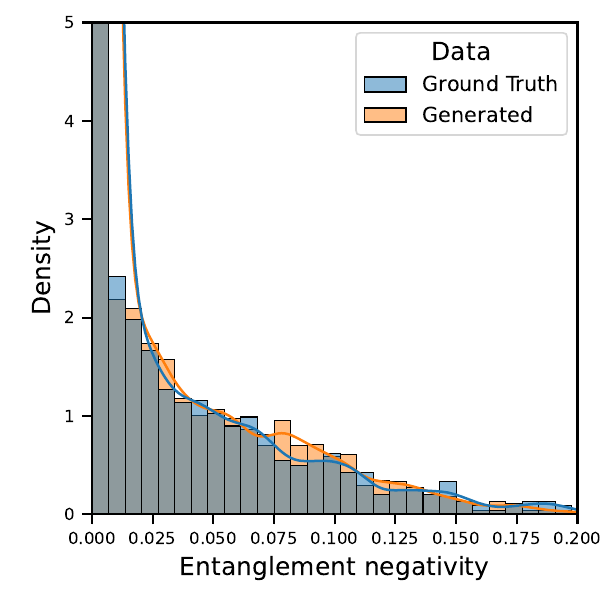}
        \end{subfigure}
        \vskip -0.2in
        \caption{Unconditional generation of the mixture of three classes of quantum states. Each figure shows an observable for comparison between unconditionally generated samples and its corresponding ground truth samples. \textbf{(1)} Leading eigenvalue versus second eigenvalue. \textbf{(2)} Real value of matrix entries in the primal space ($\mathbb{C}_{16}^{+}$). \textbf{(3)} Real value of matrix entries in the dual space ($\mathbb{C}_{16}$). \textbf{(4)} Distribution of entanglement negativity measured between qubit $1$ and the rest of the system. }
        \vskip -0.15in
    \end{figure*}

    \subsection{Summary of Methodology}
    
     Our approach can be summarized in the following steps. We take quantum state data in $\mathbb{C}_{n}^{+}$ and apply the mirror map to transform them into $\mathbb{C}_{n}$. We extract the upper triangular entries and create their real representation as a vector in the dual space $\mathbb{R}^{n^{2}}$. On the dual space, we build a diffusion model that transforms the distribution $\tilde p_{data} = ([\nabla \phi_{\#}]p_{data})$ to a Gaussian measure. We train the score neural network $s_{\theta}$ with a score-matching loss on the dual space. 
    
    For quantum state generation, we simulate the reverse-time SDE or marginally equivalent probability flow ODE with the learned score network $s_{\theta}$ to generate new samples $ Y_{\text{new}} \sim \tilde p_{data}$, which is on $\mathbb{R}^{n^{2}}$. We compute its representation $\tilde Y_{\text{new}}$ as a complex Hermitian matrices through $\tilde Y_{\text{new}} = \operatorname{Isomorphism}_{\mathbb{R}^{n^{2}} \to \mathbb{C}_{n}}(Y_{\text{new}})$, where  $\operatorname{Isomorphism}_{\mathbb{R}^{n^{2}} \to \mathbb{C}_{n}}$ stands for the isomorphic transformation between the two spaces. We generate a new quantum state by applying the inverse of mirror map, $X_{\text{new}} = \nabla \phi^{*}(\tilde Y_{\text{new}})$, which approximately satisfies the desired data distribution $p_{data}(X)$.
    
    \vspace{-0.3cm}
    \section{Experiments}
    \vspace{-0.3cm}
    As this work is the first on the generative modeling of quantum data, there is no existing approach to compare to, although both statistical and physical means for assessing the success still exist. Moreover, no public dataset has been curated either, and we will therefore prove the concept using physical but synthetic data. 
    
    More precisely, we will consider three labeled classes of data, corresponding to quantum mixed states of a $4-$qubit system with three different levels of entanglements. We will first conduct unconditional generation, from the distribution of the union of all classes. Then we will perform conditional generation, from the same union distribution but conditioned on the class label; results similar to that of individually generating from each class would be seen. Finally, we will investigate conditional generation beyond the known class labels, i.e., the ability to interpolate/extrapolate across existing classes in the training data. We defer the details of training, data preparation, and evaluation to appendix.\ref{appendix:eval},appendix.\ref{appendix:training},appendix.\ref{appendix:data-prep}.
    
    \begin{figure*}[!htb]
        \vskip -0.1in
        \centering
        \label{fig:cond}
        \begin{subfigure}
            \centering
            \includegraphics[height=1.5in]{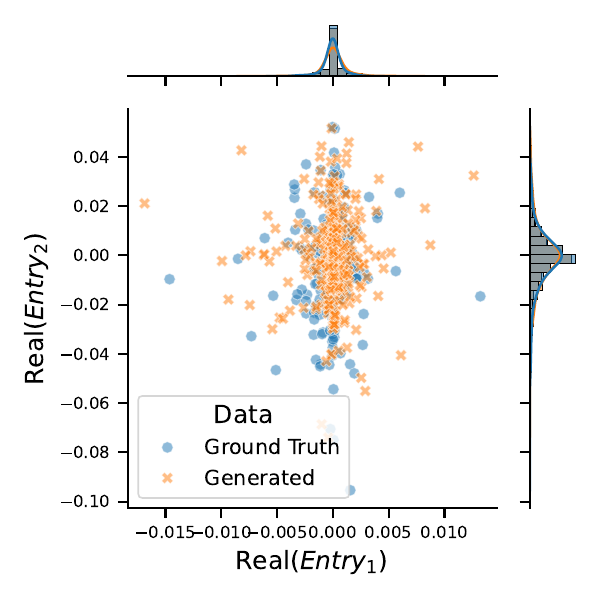}
        \end{subfigure}%
        ~ 
        \begin{subfigure}
            \centering
            \includegraphics[height=1.5in]{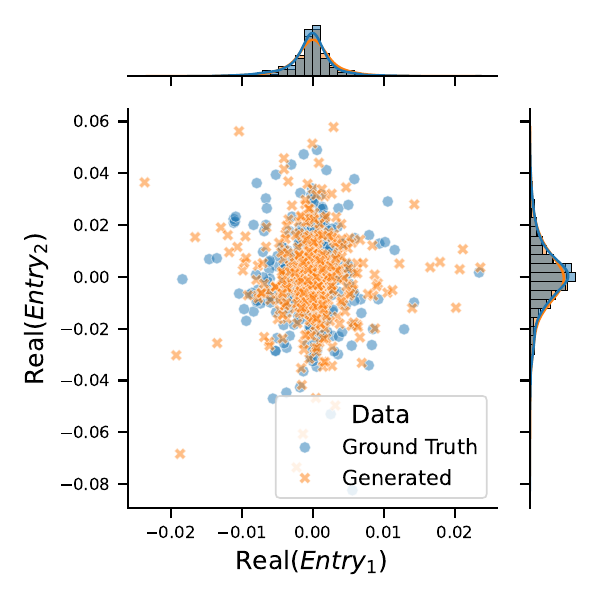}
        \end{subfigure}
        \begin{subfigure}
        \centering
        \includegraphics[height=1.5in]{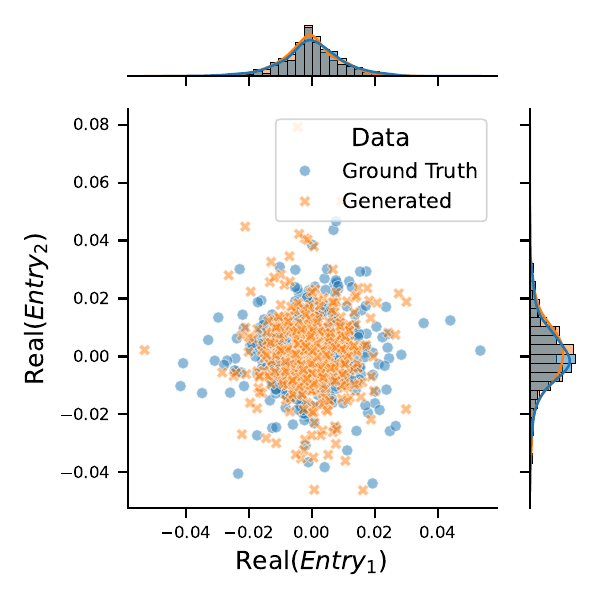}
        \end{subfigure}
        \vskip -0.1in
        \caption{Conditional generation with known labels. Each figure shows a comparison between conditionally generated samples of one class and its corresponding ground truth samples in the primal space $\mathbb{C}_{16}^{+}$. \textbf{(1)} Real value of matrix entries for product states. \textbf{(2)} Real value of matrix entries for pairwisely entangled states. \textbf{(3)} Real value of matrix entries for fully entangled states.}
        \vskip -0.2in
    \end{figure*}

    \subsection{Problem Setup}
    
    To show the efficacy of our method, we consider an example of generating quantum states based on training data consisting of quantum states with three different levels of entanglement between qubits.

    \noindent \textbf{Product state. }Let $\rho_1, \rho_2, \rho_3, \rho_4 \in \mathbb{C}_{1}^{+}$ be quantum states of one qubit. The product state with no entanglement for a $4-$qubit system is given by
    \begin{align}
        \rho_{\text{prod}} = \rho_1 \otimes \rho_2 \otimes \rho_3 \otimes \rho_4. 
    \end{align}
    To simplify the setting, we consider $\rho_{i = 1, 2, 3, 4}$ to be sampled i.i.d. from the same distribution $p_{\text{bit}}$ over quantum states of one qubit. Product state is a quantum state of the system that corresponds to the situation where there is no quantum entanglement between any of the two qubits, i.e., each qubit has a quantum state that can be independently described regardless of others. 
    
    \noindent \textbf{Pairwisely entangled state. }We now entangle qubits to create entangled state $\rho_{\text{entg}, 2}$ from product state $\rho_{\text{prod}}$. Let $U_{12}, U_{34} \in U(16)$ be complex-valued unitary matrices of dimension $16$, where $U_{12}$ corresponds to the quantum entanglement between qubit $1$ and $2$, $U_{34}$ corresponds to the quantum entanglement between qubit $3$ and $4$. A pairwisely entangled state for the $4-$qubit system is given by
    \begin{align}
    \rho_{\text{entg}, 2} = (U_{12} U_{34}) \rho_{\text{prod}} (U_{12} U_{34})^{\dagger}.
    \end{align}
    
    Such states correspond to a situation where there is simultaneous quantum entanglement between qubits $1$ and $2$, as well as qubits $3$ and $4$, but no entanglement between any other qubit pair. Since $U_{12}$ and $U_{34}$ refer to entanglement between only two qubits, the two unitary matrices are created with complex unitary matrices of dimension $4$.
    
    In this experiment, we let $M_{ij} \in \mathbb{C}^{4 \times 4}, 1\leq i < j \leq 4$ be unitary matrices of dimension $4$, sampled from the Haar measure of $U(4)$. Haar measure generalizes the concept of uniform distribution to compact Lie groups. $U_{12}, U_{34}$ are then created from $M_{12}, M_{34}$ in the following way:
    \begin{align*}
    U_{12} = M_{12} \otimes \mathbb{I}_{4\times 4}, \quad U_{34} =  \mathbb{I}_{4\times 4} \otimes M_{34}.
    \end{align*}
    
    \begin{figure*}[!htb]
        \begin{subfigure}
        \centering
        \includegraphics[width=\textwidth]{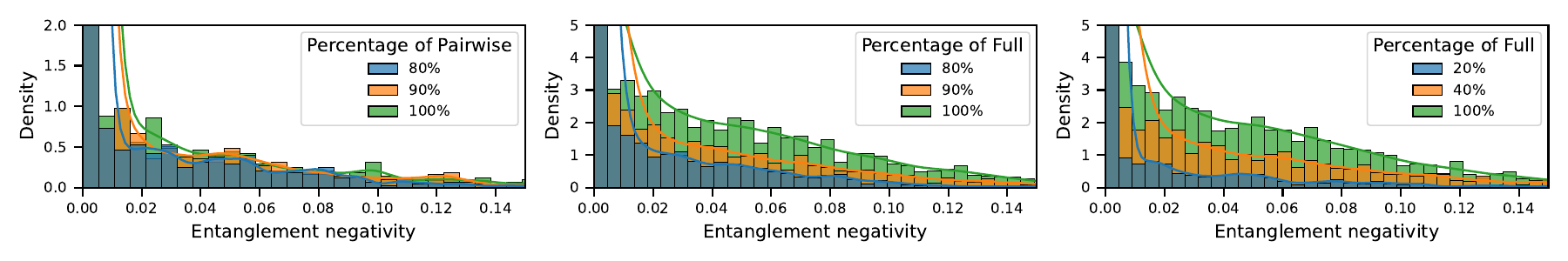}
        \end{subfigure}
        \vskip -0.3in
        \caption{Conditional generation with unseen labels.  Each figure represents the distribution of the entanglement negativity of samples generated with unseen labels that are convex combinations of one-hot encoding of seen labels. Entanglement negativity is measured between qubit $1$ and the rest of the system. \textbf{(Left)} Interpolation between product state and pairwisely entangled state. \textbf{(Middle)} Interpolation between pairwisely entangled states and fully entangled states. \textbf{(Right)} Interpolation between product states and fully entangled states.}
        \label{fig:interp}
    \end{figure*}
    
    \noindent \textbf{Fully entangled state.} We now entangle more qubits to create further entangled state $\rho_{\text{entg}, \text{all}}$ from product state $\rho_{\text{prod}}$. 
    For $1 \leq i < j \leq 4$, let $U_{ij} \in U(16)$ be complex unitary matrices of dimension $16$, where $U_{ij}$ corresponds to the quantum entanglement between qubit $i$ and $j$. Consider a fully entangled state for the $4-$qubit system given by
    \vspace{-0.2em}
    \begin{align}
        \rho_{\text{entg}, \text{all}} = (\prod_{i < j} U_{ij}) \rho_{\text{prod}} (\prod_{i < j} U_{ij})^{\dagger}.
    \end{align}
    Fully entangled state corresponds to a situation where there is simultaneous entanglement between any pair of qubits. Each matrix $U_{ij}$ corresponds to the entanglement between only two qubits, thus the matrix is created using complex unitary matrices of dimension $4$.
    
    $U_{ij}$ are created from $M_{ij}$ in the following way:
    \begin{align*}
        & U_{12} = M_{12} \otimes \mathbb{I}_{4\times 4}, 
        && U_{13} = \operatorname{Perm}_{23}(M_{13} \otimes \mathbb{I}_{4\times 4}), 
        && U_{14} = \operatorname{Perm}_{13}(\mathbb{I}_{4\times 4} \otimes M_{14}),  \\
        & U_{23} = \mathbb{I}_{2\times 2} \otimes M_{23} \otimes \mathbb{I}_{2\times 2},
        && U_{24} =  \operatorname{Perm}_{23}(\mathbb{I}_{4\times 4} \otimes M_{24}), 
        && U_{34} = \mathbb{I}_{4\times 4} \otimes M_{34},
    \end{align*}
    where $\operatorname{Perm}_{ij}$ permutes rows and columns corresponding to qubit $i$ and $j$, $\mathbb{I}_{n\times n}$ is the identity matrix.

    \vspace{-0.2cm}
    \subsection{Results}
    \vspace{-0.2cm}
    \noindent \textbf{SPDM accurately learns the distribution of quantum states.} Table \ref{tb:distributional_metric} summarizes our quantitative results for both conditional and unconditional generation of three classes of quantum states. Figure \ref{fig:uncond} and Figure \ref{fig:cond} depict the generated samples against ground truth. Our method manages to learn accurately the correct distribution of eigenvalues, which is extremely sensitive to all matrix entries. Note that product states, pairwisely entangled states, and fully entangled states considered in this work share an identical eigenvalue distribution since matrix spectrum is invariant under unitary conjugations. However, they have distinct marginal distributions for each entry due to the quantum entanglement. Our method successfully predicts the joint distribution of all the entries in the density matrix, which is also confirmed by small values in all distribution-based metrics listed in Table \ref{tb:distributional_metric}. 

        \begin{table*}[h]
        \centering
        \begin{tabular}{ lcccccc }
        & $\textbf{Class}$ & $\textbf{SWD}\downarrow$ & $\textbf{MSWD} \downarrow $ & $\mathbf{W}_1\downarrow $ & $\textbf{Energy-MMD}\downarrow $ & $\textbf{Negativity} (\mathbf{W}_1)\downarrow $  \\
        \hline
        \textbf{Uncond.} & Mixture & $0.00054$  & $0.00094$&  0.020&0.000076& 0.0000110\\ 
        
        \hline
        & Product & $0.00052$  & 0.00079&  0.011& 0.000070& 0.0000025\\ 
        \textbf{ Cond.} & Pairwisely & $0.00064$ &0.00100 & 0.018 &0.000082& 0.0000061 \\ 
        & Fully & $0.00056$  & 0.00092 & 0.026&0.000100& 0.0000062 \\ 
        \hline
        \textbf{Mag. ref.} & -- & 0.0854 &0.1554 &1.7484 &1.1236 & 0.0134 \\
        \hline
        \end{tabular}
        \caption{Unconditional and conditional generation of quantum states. We report metrics computed using $3000$ generated samples and ground truth samples. \{Sliced Wasserstein Distance (SWD), Maximum Sliced Wasserstein Distance (MSWD), Energy Maximum Mean Discrepancy (Energy-MMD), 1-Wasserstein distance ($W_{1}$)\} 
        are computed using density matrices of generated quantum state. Negativity ($W_{1}$) is computed by measuring the $W_{1}$ distance between the distribution of entanglement negativity. More details on evaluation metrics can be found in Appendix \ref{appendix:eval}. The last row is just providing a reference for interpreting how small values mean close distributions, as it compares the training set against randomly generated density matrices ($16\times16$ Hermitian with eigenvalues uniformly sampled from $[0,1]$ and eigenspace matrix sampled from Haar on $U(16)$), for which metrics values should be large.} 
        \label{tb:distributional_metric}
        \end{table*}

    \noindent \textbf{SPDM generates the correct level of quantum entanglement between qubits.} Other than eigenvalues and distribution-based metrics, another assessment of generation quality is whether the generated samples demonstrate the same level of entanglement as the training data. We use entanglement negativity to capture the degree of quantum entanglement between two subsystems. As is evident from Table \ref{tb:distributional_metric} and Figure \ref{fig:uncond}, our method accurately generates samples with the same amount of quantum entanglement as the training data. This implies that under our framework, diffusion models manage to learn the sophisticated relationships between entangled entries and reproduce such entanglements in the generation process. 
    
    \noindent \textbf{SPDM enables the design of physically-meaningful new quantum states.} We generate new labels by taking a convex combination of the one-hot encodings of seen classes. Then we conditionally generate samples that correspond to these new labels with our model. Figure \ref{fig:interp} demonstrated the quantum entanglement level of the resulting data. When generating under labels that are mixtures of two classes with different entanglements, the model produces data with an entanglement level in the middle ground of the two classes, which is in accordance with the interpolation ratio we used in generating new labels. This implies that our method manages to learn a meaningful embedding of class labels that also reflects the interpolation of entanglement level among quantum data. The model allows an extrapolation to unseen quantum states. This means that our model is capable of designing quantum states with specific (new) entanglement level controlled by a manually created label.
    \vspace{-0.2cm}
    \section{Conclusion, Limitation, and Future Work}
    \vspace{-0.2cm}
    We developed the first diffusion-based method for generative modeling quantum states. It hard codes physical knowledge into the learning models and generates samples that strictly satisfy the structural constraints of quantum states, resulting in accurate generation represented by
    density matrices. It can even create physically-meaningful states of unseen level of entanglements, via conditional generation. 

    In the future, it will be interesting  to design neural network architectures for parametrizing the score function that specialize in quantum state generation. SOTA scalability is already demonstrated, but to further improve, generic score parametrization may be insufficient. The training procedure has also not been the focus, but there is also significant potential for improvement (e.g., \citep{karras2022elucidating}). Finally, this work mainly focuses on learning methodology development, and physical applications to real quantum problems will also be left as an important future direction.

\newpage

\bibliographystyle{abbrv}
\bibliography{main}

\newpage

\appendix

\section{Appendix}
\subsection{Vanilla SGM on Quantum State Dataset}
\begin{figure}[h]
    \includegraphics[width=\textwidth]{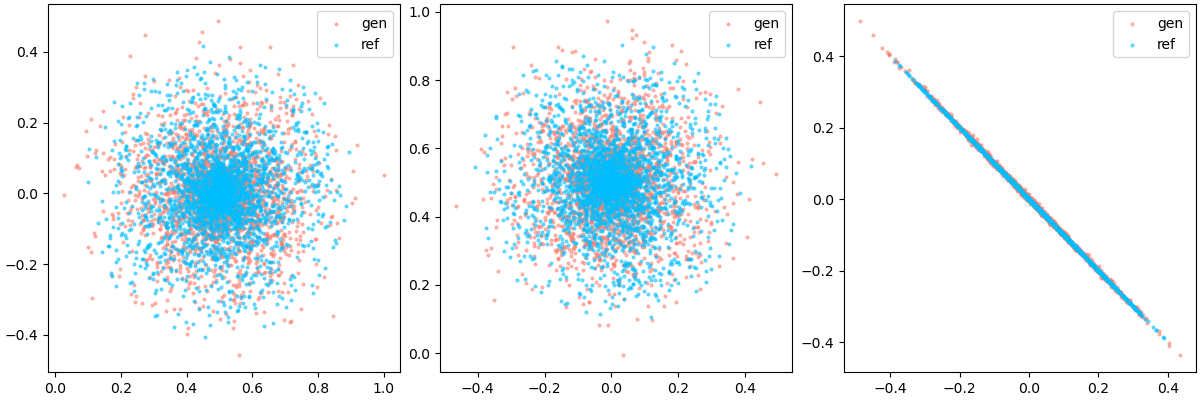}
    \caption{SGM can easily recover the distribution of 1 qubit quantum state.}
    \label{fig:sgm_success}
    \vskip -0.05in
\end{figure}

\begin{figure}[h]
    \includegraphics[width=\textwidth]{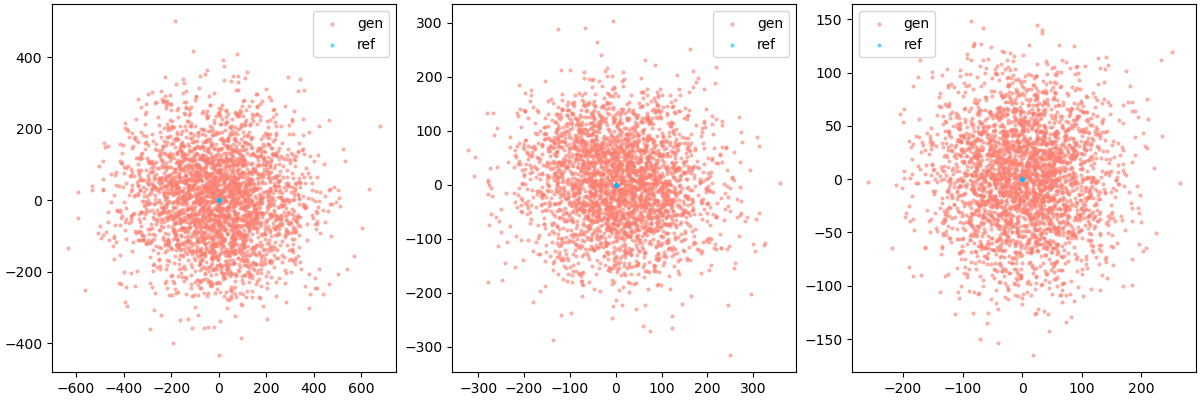}
    \caption{When the dimension increases, SGM cannot easily recover the distribution of the 4 qubit quantum state due to the complex data structure.}
    \label{fig:sgm_fail}
    \vskip -0.05in
\end{figure}

In our research, we demonstrate that conventional score-based generative models (SGMs) fall short in reliably generating high-dimensional quantum datasets. Figure \ref{fig:sgm_success} exemplifies that, although these SGMs can precisely mimic the distribution of a 1-qubit quantum state (4 dimensions), they face significant challenges due to the curse of dimensionality when the dimensionality is expanded to a 4-qubit state (256 dimensions), as highlighted in Figure \ref{fig:sgm_fail}. Note that in Figure \ref{fig:sgm_fail}, the compared distribution is the true data distribution (in blue) with the generated data distribution (in red). While the blue points seem to be a single point dirac distribution, in fact it has a similar structure as the true distribution depicted in Figure \ref{fig:sgm_success}. The true distribution in Figure \ref{fig:sgm_fail} looks like a single point due to the drastic failure of Euclidean SGM on this task and an exploded data scale of the generated samples. This limitation likely stems from the current neural network parameterization strategies. Identifying an efficient network architecture capable of effectively capturing the essence of quantum states remains a pivotal, yet unresolved, question, which we aim to address in future work.

\subsection{Proof of Proposition.\ref{prop:mirror_fn}}\label{Appendix:prop:mirror_fn_proof}
\begin{proof}
We first show that $\nabla \phi$ is a mapping from $\mathbb{C}_{n}^{+}$ to $\mathbb{C}_n$ and $\nabla \phi^*$ is a mapping from $\mathbb{C}_n$ to $\mathbb{C}_{n}^{+}$.
For any $X \in \mathbb{C}_{n}^{+}$, $Y \in \mathbb{C}_n$, we denote their spectral decomposition as $X = \sum_{i} \lambda_i q_{i} q_{i}^{\dagger}$, $Y = \sum_i w_i p_i p_i^{\dagger}$ with complex-valued eigenvectors $\{q_i\}$ and $\{p_i \}$ form orthonormal basis of $\mathbb{C}^{n}$, and eigenvalues $\lambda_i > 0$, $w_i \in \mathbb{R}$,
\begin{align*}
    \nabla \phi(X) & = I + \log X = I + \sum_i (\log \lambda_i) q_i q_i^{\dagger} \\
    (\nabla \phi(X))^{\dagger} & = I^{\dagger} + \sum_{i} \overline{\log \lambda_i} (q_i q_i^{\dagger})^{\dagger} = I + \sum_{i} (\log \lambda_i) q_{i} q_{i}^{\dagger} \\
    & = \nabla \phi(X)
\end{align*}
This implies that $\nabla \phi(X) \in \mathbb{C}_n$ is a complex Hermitian matrix for any $X \in \mathbb{C}_{n}^{+}$. Similarly,
\begin{align*}
\nabla \phi^*(Y) & = \exp(Y - I) = \sum_{i} \exp(w_i - 1) p_i p_i^{\dagger} \\
    (\nabla \phi^*(Y))^{\dagger} & = \sum_i \overline{\exp(w_i - 1)} (p_i p_i^{\dagger})^{\dagger} = \nabla \phi^*(Y)
\end{align*}
Since $\nabla \phi^* (Y)$ has positive eigenvalues $\exp(w_i - 1)$, $\nabla \phi^*(Y) \in \mathbb{C}_{n}^{+}$ for any $Y \in \mathbb{C}_n$.

Now we need to show that $\nabla \phi$ and $\nabla \phi^*$ are inverse of each other.
\begin{align*}
    \nabla \phi^{*}(\nabla \phi(X)) & = \exp(I + \sum_i \log \lambda_i q_i q_i^{\dagger} - I) \\
    & = \sum_{i} \lambda_i q_i q_i^{\dagger} = X
\end{align*}
This implies that $\nabla \phi^* \circ \nabla \phi$ is the identity map on $\mathbb{C}_{n}^{+}$.
\begin{align*}
    \nabla \phi(\nabla \phi^* (Y)) & = \nabla \phi (\exp(\sum_{i} (w_i - 1) p_i p_i^{\dagger})) \\
    & = I + \log(\sum_{i}\exp(w_i - 1) p_i p_i^{\dagger}) \\
    & = I + \sum_{i} (w_i - 1) p_i p_i^{\dagger} = Y
\end{align*}
Similarly, $\nabla \phi \circ \nabla \phi^*$ is the identity map on $\mathbb{C}_n$. Therefore, we conclude that $(\nabla \phi, \nabla \phi^*)$ is a pair of bijective mappings that are inverse of each other. 
\end{proof}

\subsection{Training and neural network} \label{appendix:training}
We train on a dataset of $900,000$ samples, distributed equally across all three classes of quantum states. The entire dataset is first transformed to the dual space by applying the mirror map. Then a DDPM is trained on the transformed dataset. We train for $1,000,000$ iterations with AdamW optimizer, using an initial learning rate of $10^{-3}$ and decay rate of $0.995$ every $1000$ AdamW steps.

Our neural network takes an input pair $(x,t, c)$, where $x$ is the spatial input, $t$ is the time, and $c$ is the condition. Time $t$ is passed through a module to generate a standard sinusoidal embedding and then fed to $2$ fully connected layers with Sigmoid Linear Unit (SiLU) and generate an output $t_{\text{out}}$. Spatial input $x$ is passed through an MLP with $8$ residual blocks, each containing $4$ linear layers with hidden dimension $512$ and SiLU activation. This generates an output $y_{\text{out}}$. The condition $c$, considered to be a one-hot encoding in our case, is fed to $3$ fully connected layers with SiLU and generates an output $c_{\text{out}}$. Our final output $\text{out}$ is computed through,
\begin{align*}
\text{out} = \text{Outmod}(\operatorname{GroupNorm}(y_{\text{out}} + t_{\text{out}} + c_{\text{out}}))
\end{align*}
where $\text{Outmod}$ is an out module that consists of $4$ fully connected layers with hidden dimension $512$ and SiLU activation, $\operatorname{GroupNorm}$ stands for group normalization.

All the experiments are performed on one RTX 4090 and one RTX 3090.

\subsection{Data Preparation}\label{appendix:data-prep}
    This section describes how we generate thethree classes of quantum states, which altogether constitute our training set. 
    
    \noindent \textbf{Single qubit quantum state distribution.} In the generation process of product state, we consider the scenario where each qubit has a quantum state sampled i.i.d. from the same distribution $p_{\text{bit}}$, which is defined by the following procedure. 
    Let $Y = Q \Lambda Q^{\dagger}$, where $Q \sim \operatorname{Haar}(U(2))$, $\Lambda = \operatorname{diag}([\lambda_1, \lambda_2])$ with $\lambda_1, \lambda_2 \sim \operatorname{Uniform}[\lambda_{\text{min}}, \lambda_{\text{max}}]$. 
    $X \in \mathbb{C}_{1}^{+}$ is computed as
    \begin{align}
    \label{eq:one-qubit-distribution}
        X = \frac{1}{\operatorname{Tr}(Y)} Y.
    \end{align}
    We consider $p_{\text{bit}}$ to be the distribution of $X$ generated via eq.\eqref{eq:one-qubit-distribution}. By construction, $X$ is a positive definite Hermitian matrix with trace $1$, thus a valid quantum state.

    \noindent \textbf{Haar measure of $U(n)$.} To generate the training data for pairwisely entangled and fully entangled states, we chose to sample from the Haar measure on $U(4)$ to create unbiased quantum entanglement matrices $U_{ij}$ between qubit $i$ and $j$. $U(n)$ is the set of matrices defined as,
    \begin{align*}
        U(n) = \{M \in \mathbb{C}^{n \times n}, M^{\dagger}M = MM^{\dagger} = \mathbb{I}_{n} \}.
    \end{align*}
    Sampling a probability distribution on $U(n)$, such as the Haar measure, is also an interesting task. One may view it as a sampling problem on a constrained set, but note $U(n)$ is actually a non-convex constrained set, and thus common constrained sampling tools (e.g., projected Langevin \citep{bubeck2015finite}, sophisticated MCMC walkers \citep{lovasz1993random, lovasz1999hit, kannan2009random, chen2018fast}, mirror Langevin \citep{zhang2020wasserstein, li2022mirror}) may not suit well. Fortunately, due to the Lie group structure of $U(n)$, the problem can be solved by considering the dynamics,
    \begin{equation}
    \label{eq:lie-dynamic}
        \begin{cases}
            \dot {g} = g \xi, \\
            \mathrm{d} \xi = - \gamma \xi \mathrm{d}t + \sqrt{2 \gamma} \mathrm{d} W_{\mathfrak{g}},
        \end{cases}
    \end{equation}
    where $g \in U(n)$, $\xi \in \mathfrak{g} = \mathfrak{u}(n)$ is an element in the Lie algebra, $W_{\mathfrak{g}}$ is the Brownian motion over the vector space $\mathfrak{u}(n)$. The Lie algebra and Brownian motion is defined as,
    \vskip -0.2in
    \begin{align*}
    \mathfrak{u}(n) = \{M \in \mathbb{C}^{n \times n}, M^{\dagger} = - M \}, \mathrm{d}W_{\mathfrak{g}} = \sum_{i = 1}^{n^{2}} \mathrm{d}W_{i} \cdot e_{i}
    \end{align*}
    \vskip -0.2in
    where $\{e_1, \dots, e_{n^{2}} \} \text{ spans } \mathfrak{u}(n)$, $\mathrm{d}W_1, \dots, \mathrm{d}W_{n^{2}}$ are independent one dimensional Brownian motion. One notable property of the diffusion process described in \eqref{eq:lie-dynamic} is that it converges to a unique invariant distribution and the $g$-marginal of its invariant distribution is the Haar measure of $U(n)$ \citep{kong2024convergence}. Therefore, by numerically integrating \eqref{eq:lie-dynamic}, we can approximately sample uniformly from the unitary group $U(4)$ to generate the quantum entanglement matrices. This numerical integration can be done computational-efficiently with $g$ exactly remaining on $U(4)$ via the operator splitting approach \citep{tao2020variational}.

   \subsection{Evaluation}\label{appendix:eval}
    
    In this section, we discuss the metrics and observables (since the data distribution is high dimensional) used for evaluating the sample quality of the generated quantum states. Both statistical and physical quantities will be used.
    
    \noindent \textbf{Entanglement negativity.} Entanglement negativity is a non-negative measure of the amount of quantum entanglement present in a quantum state. Entanglement negativity is an entanglement monotone 
    whose value does not increase under local operations and classical communication. Consider a composite quantum system consisting of two subsystems $A$ and $B$, described by a density matrix $\rho$ in a Hilbert space that is the tensor product of the spaces of $A$ and $B$. 
    We denote the transpose operator on the space of subsystem A as $T_{A}$. We denote the partial transpose of $\rho$ with respect to subsystem $A$ as $\rho^{\Gamma_{A}}$, which can be computed as $\rho^{\Gamma_{A}} = (T_{A} \otimes I) \rho$.
    Then, the entanglement negativity $\mathcal{N}(\rho)$ is defined as,
    \begin{align*}
    \mathcal{N}(\rho) = \Big|\sum_{\lambda_i < 0} \lambda_i \Big | = \sum_{i} \frac{|\lambda_i| - \lambda_i}{2}
    \end{align*}
    where $\lambda_i$ are eigenvalues of $\rho^{\Gamma_{A}}$. A high value of $\mathcal{N}(\rho)$ represents a high level of quantum entanglement.
    
    \noindent \textbf{Distributional-wise metric.} To evaluate the similarity between high-dim probability distributions, we mainly use Sliced Wasserstein Distance (SWD) \citep{kolouri2019generalized}, Maximum Sliced Wasserstein Distance (MSWD) \citep{deshpande2019max}, and Maximum Mean Discrepancies (MMD) \citep{gretton2012kernel}, in addition to the standard Wasserstein distance, as they suffer less from the curse of dimensionality. Within the family of MMDs, we consider Energy-MMD, which is computed with kernel $k(x,y) = -\|x-y\|_2$. We also compute the 1-Wasserstein distance based on values of the physically meaningful low-dim observable of entanglement negativity.

\begin{remark}
The 1-Wasserstein Distance suffers from the curse of dimensionality seriously which is pointed out in Appendix.C in \citep{chen2023deep}. Therefore, we use various metrics to comprehensively evaluate the distance between high dimensional distributions. For the scalar distribution of negativity, we stick to using 1-Wasserstein distance. In the main results, when evaluating the sample quality of generated quantum states (which is a density of $512$ dimensions), we use the Sliced-Wasserstein Distance (SWD) type and Maximum Mean Discrepancies (MMD) type of metrics as our major criterion. These metrics include Energy Maximum Mean Discrepancy (Energy-MMD), Max-sliced Wasserstein distance (MSWD), and Sliced-Wasserstein Distance (SWD). Here, Energy-MMD is the MMD distance computed with kernel $k(x,y) = -\|x-y\|_2$. Our implementation is adapted from \href{https://www.kernel-operations.io/geomloss/}{Geoloss ($W_1$, Energy-MMD)} and \href{https://pythonot.github.io/gen_modules/ot.sliced.html}{POT (SWD, MSWD)}. 
\end{remark}

\subsection{Additional Numerical Experiments} \label{append:addtional}
In this section, we provide additional plots for experiments on $6-$ qubits. The description of these results can be found in the caption of each figure.

\begin{figure}[h]
    \centering
    \begin{subfigure}
        \centering
        \includegraphics[height=2in]{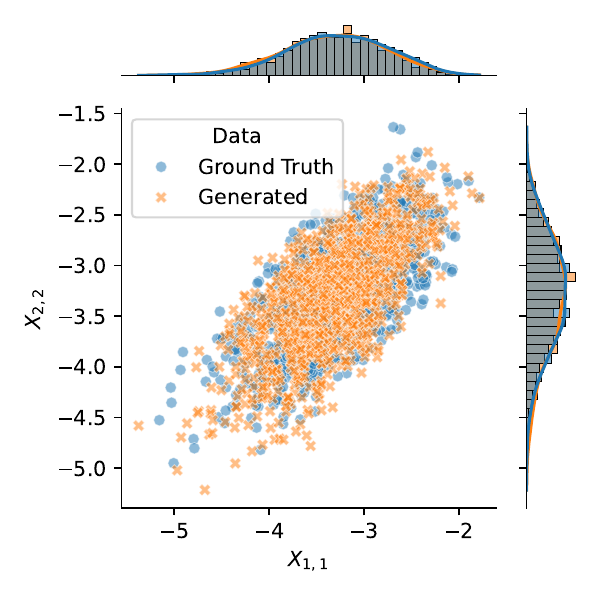}
    \end{subfigure}%
    ~ 
    \begin{subfigure}
        \centering
        \includegraphics[height=2in]{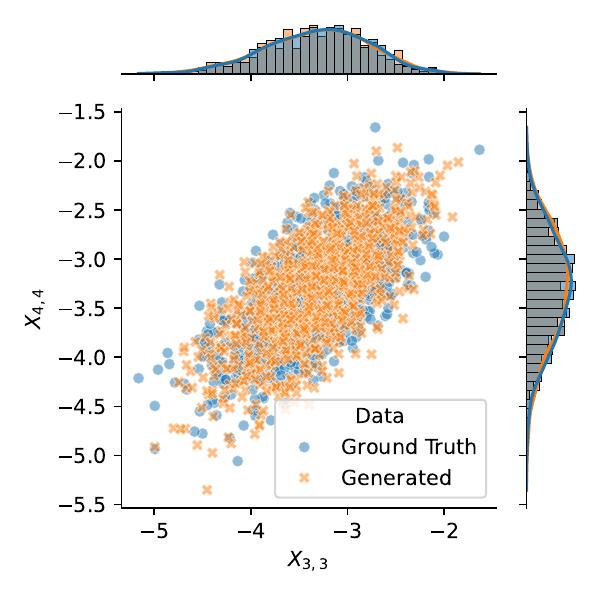}
    \end{subfigure}
    \begin{subfigure}
    \centering
    \includegraphics[height=2in]{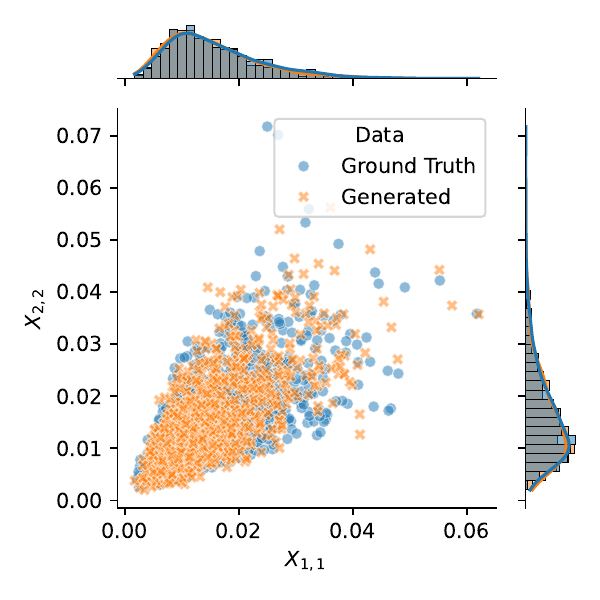}
    \end{subfigure}
    \begin{subfigure}
    \centering
    \includegraphics[height=2in]{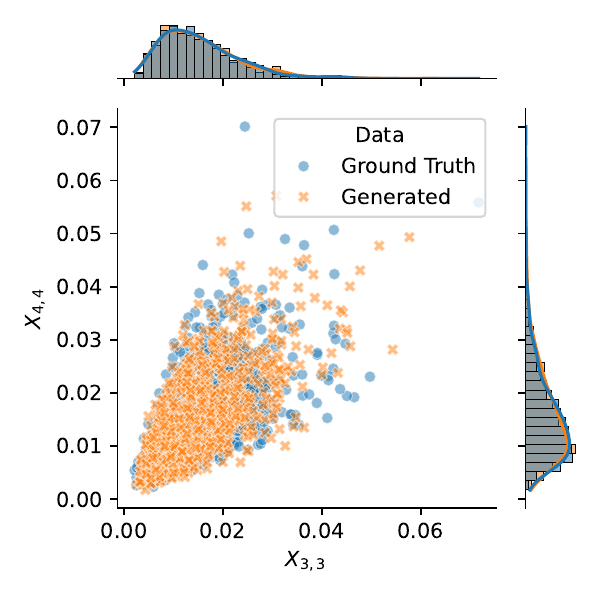}
    \end{subfigure}
    \vskip -0.2in
    \caption{\textbf{Unconditional generation of product states with $6$ qubits}. Each qubit is i.i.d. sampled with eigenvalues bounded between $1$ and $3$. Each figure shows an observable for comparison between unconditionally generated samples and their corresponding ground truth samples. From Left to Right: 
    \textbf{(1)} Real value of matrix entries in the dual space ($\operatorname{Real}(X_{1,1})$ v.s. $\operatorname{Real}(X_{2,2}))$
    \textbf{(2)} Real value of matrix entries in the dual space ($\operatorname{Real}(X_{3,3})$ v.s. $\operatorname{Real}(X_{4,4}))$ 
    \textbf{(3)} Real value of matrix entries in the primal space ($\operatorname{Real}(X_{1,1})$ v.s. $\operatorname{Real}(X_{2,2}))$
    \textbf{(4)} Real value of matrix entries in the primal space ($\operatorname{Real}(X_{3,3})$ v.s. $\operatorname{Real}(X_{4,4}))$}
    \vskip -0.15in
\end{figure}

\newpage

\begin{figure}[h]
    \centering
    \begin{subfigure}
        \centering
        \includegraphics[height=2in]{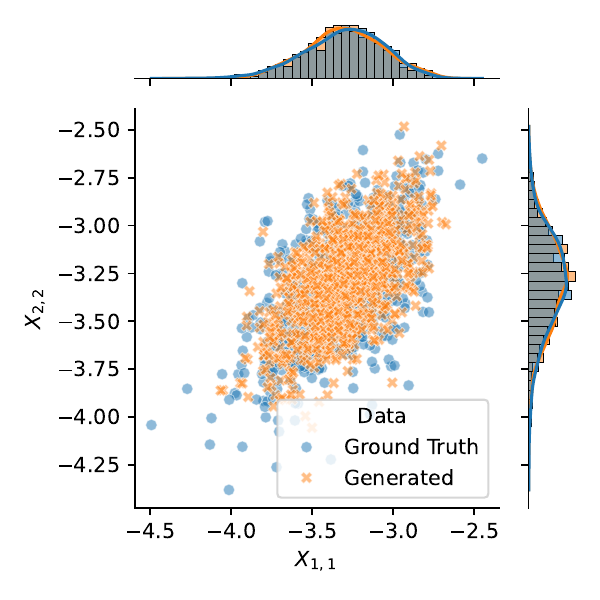}
    \end{subfigure}%
    ~ 
    \begin{subfigure}
        \centering
        \includegraphics[height=2in]{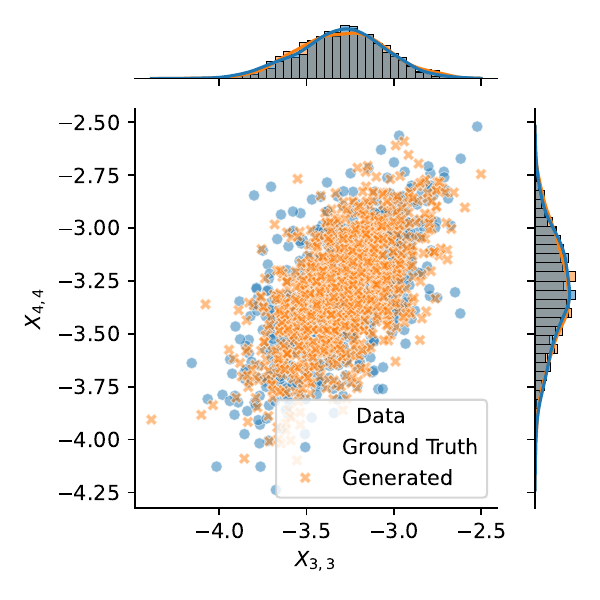}
    \end{subfigure}
    \begin{subfigure}
    \centering
    \includegraphics[height=2in]{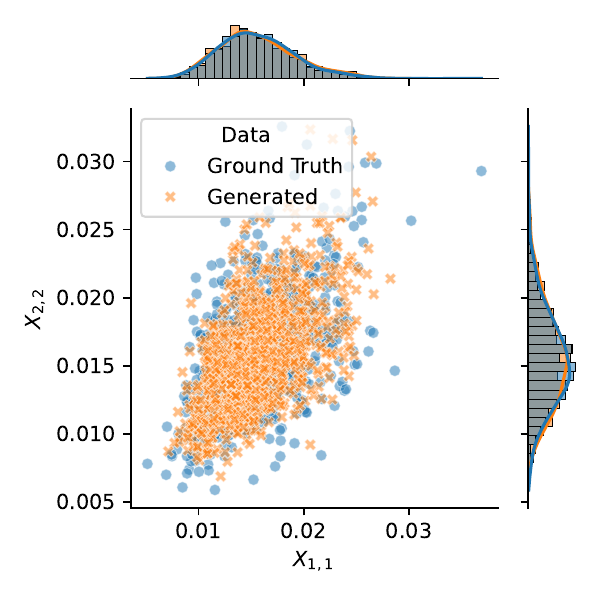}
    \end{subfigure}
    \begin{subfigure}
    \centering
    \includegraphics[height=2in]{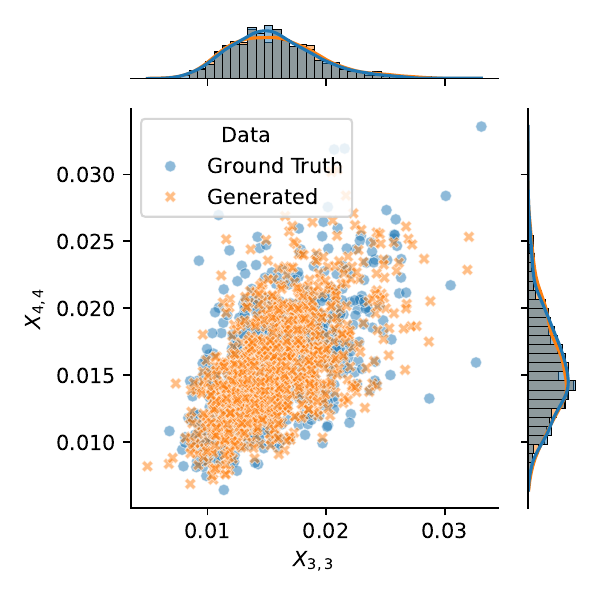}
    \end{subfigure}
    \vskip -0.2in
    \caption{\textbf{Unconditional generation of pairwisely-entangled states with $6$ qubits}. The product states are generated as tensor products of $6$ single qubit quantum state density matrix. Each qubit is i.i.d. sampled with eigenvalues bounded between $1$ and $3$. The pairwisely entangled states are generated from the product states by imposing entanglement on qubit pairs $\{1,2\}, \{3,4\}, \{5,6\}$. The entanglement matrices are sampled from the Haar measure of $U(4)$. Each figure shows an observable for comparison between unconditionally generated samples and their corresponding ground truth samples. From Left to Right: 
    \textbf{(1)} Real value of matrix entries in the dual space ($\operatorname{Real}(X_{1,1})$ v.s. $\operatorname{Real}(X_{2,2}))$
    \textbf{(2)} Real value of matrix entries in the dual space ($\operatorname{Real}(X_{3,3})$ v.s. $\operatorname{Real}(X_{4,4}))$ 
    \textbf{(3)} Real value of matrix entries in the primal space ($\operatorname{Real}(X_{1,1})$ v.s. $\operatorname{Real}(X_{2,2}))$
    \textbf{(4)} Real value of matrix entries in the primal space ($\operatorname{Real}(X_{3,3})$ v.s. $\operatorname{Real}(X_{4,4}))$}
    \vskip -0.15in
\end{figure}

\newpage
\begin{figure}[h]
    \centering
    \begin{subfigure}
        \centering
        \includegraphics[height=2in]{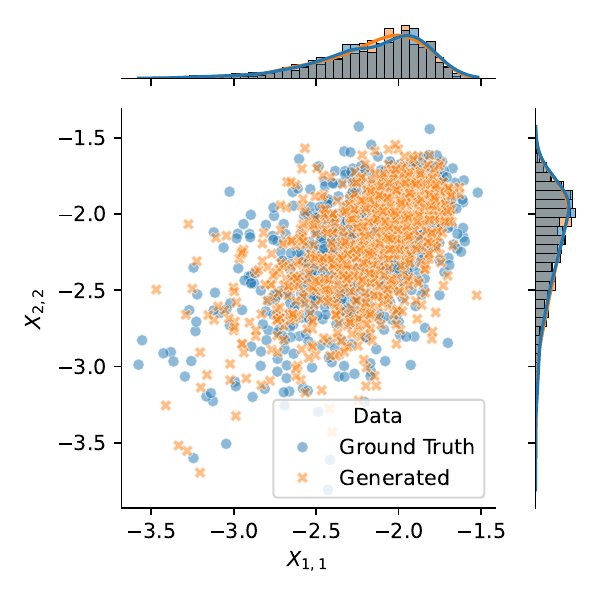}
    \end{subfigure}%
    ~ 
    \begin{subfigure}
        \centering
        \includegraphics[height=2in]{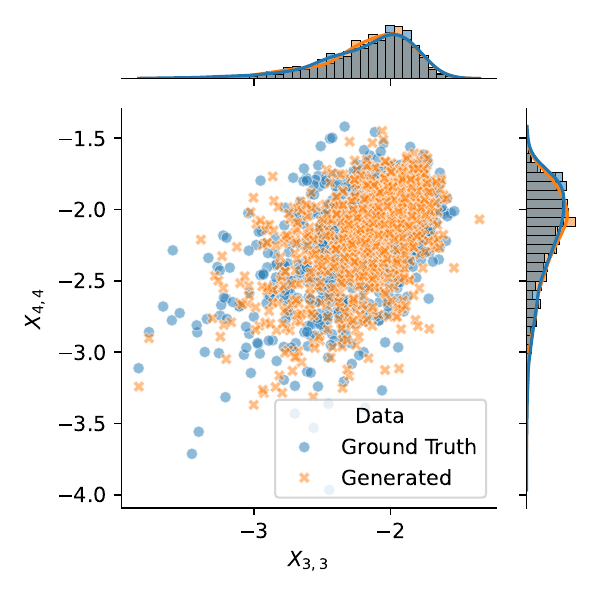}
    \end{subfigure}
    \begin{subfigure}
    \centering
    \includegraphics[height=2in]{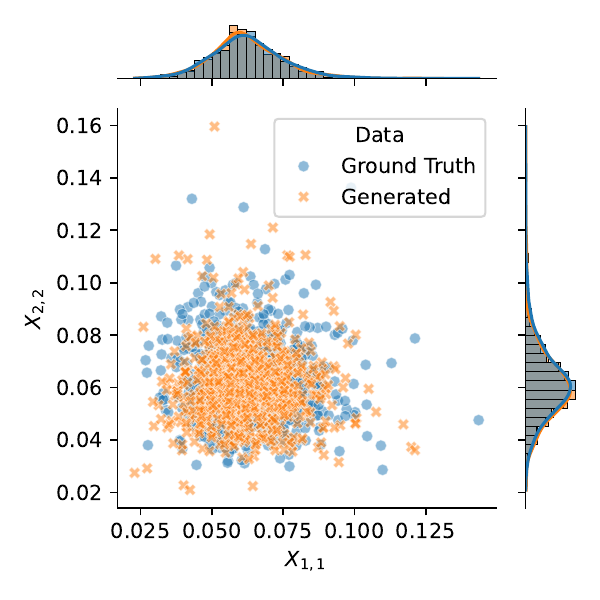}
    \end{subfigure}
    \begin{subfigure}
    \centering
    \includegraphics[height=2in]{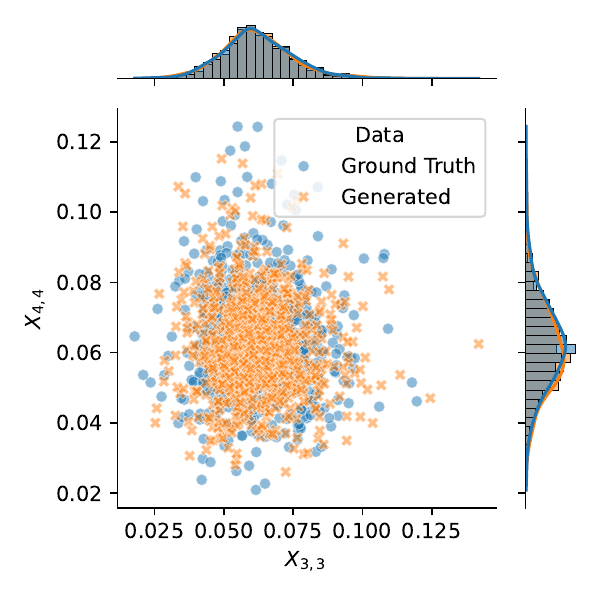}
    \end{subfigure}
    \vskip -0.2in
    \caption{\textbf{Unconditional generation of maximally-entangled quantum states with $4$ qubits}. Entanglement matrix $U$ are sampled from Haar measure of $U(16)$ instead of pairwisely from all pairs of qubits. Each figure shows an observable for comparison between unconditionally generated samples and their corresponding ground truth samples. From Left to Right: 
    \textbf{(1)} Real value of matrix entries in the dual space ($\operatorname{Real}(X_{1,1})$ v.s. $\operatorname{Real}(X_{2,2}))$
    \textbf{(2)} Real value of matrix entries in the dual space ($\operatorname{Real}(X_{3,3})$ v.s. $\operatorname{Real}(X_{4,4}))$ 
    \textbf{(3)} Real value of matrix entries in the primal space ($\operatorname{Real}(X_{1,1})$ v.s. $\operatorname{Real}(X_{2,2}))$
    \textbf{(4)} Real value of matrix entries in the primal space ($\operatorname{Real}(X_{3,3})$ v.s. $\operatorname{Real}(X_{4,4}))$}
    \vskip -0.15in
\end{figure}

\newpage
\section*{NeurIPS Paper Checklist}
\begin{enumerate}

\item {\bf Claims}
    \item[] Question: Do the main claims made in the abstract and introduction accurately reflect the paper's contributions and scope?
    \item[] Answer: \answerYes{} 
    \item[] Justification: we provided the scope and contribution in the abstract.
    \item[] Guidelines:
    \begin{itemize}
        \item The answer NA means that the abstract and introduction do not include the claims made in the paper.
        \item The abstract and/or introduction should clearly state the claims made, including the contributions made in the paper and important assumptions and limitations. A No or NA answer to this question will not be perceived well by the reviewers. 
        \item The claims made should match theoretical and experimental results, and reflect how much the results can be expected to generalize to other settings. 
        \item It is fine to include aspirational goals as motivation as long as it is clear that these goals are not attained by the paper. 
    \end{itemize}

\item {\bf Limitations}
    \item[] Question: Does the paper discuss the limitations of the work performed by the authors?
    \item[] Answer: \answerYes{} 
    \item[] Justification: please see the Conclusion section in the main text.
    \item[] Guidelines:
    \begin{itemize}
        \item The answer NA means that the paper has no limitation while the answer No means that the paper has limitations, but those are not discussed in the paper. 
        \item The authors are encouraged to create a separate "Limitations" section in their paper.
        \item The paper should point out any strong assumptions and how robust the results are to violations of these assumptions (e.g., independence assumptions, noiseless settings, model well-specification, asymptotic approximations only holding locally). The authors should reflect on how these assumptions might be violated in practice and what the implications would be.
        \item The authors should reflect on the scope of the claims made, e.g., if the approach was only tested on a few datasets or with a few runs. In general, empirical results often depend on implicit assumptions, which should be articulated.
        \item The authors should reflect on the factors that influence the performance of the approach. For example, a facial recognition algorithm may perform poorly when image resolution is low or images are taken in low lighting. Or a speech-to-text system might not be used reliably to provide closed captions for online lectures because it fails to handle technical jargon.
        \item The authors should discuss the computational efficiency of the proposed algorithms and how they scale with dataset size.
        \item If applicable, the authors should discuss possible limitations of their approach to address problems of privacy and fairness.
        \item While the authors might fear that complete honesty about limitations might be used by reviewers as grounds for rejection, a worse outcome might be that reviewers discover limitations that aren't acknowledged in the paper. The authors should use their best judgment and recognize that individual actions in favor of transparency play an important role in developing norms that preserve the integrity of the community. Reviewers will be specifically instructed to not penalize honesty concerning limitations.
    \end{itemize}

\item {\bf Theory Assumptions and Proofs}
    \item[] Question: For each theoretical result, does the paper provide the full set of assumptions and a complete (and correct) proof?
    \item[] Answer: \answerYes{} 
    \item[] Justification: We provide the full proof and the certain assumptions.
    \item[] Guidelines:
    \begin{itemize}
        \item The answer NA means that the paper does not include theoretical results. 
        \item All the theorems, formulas, and proofs in the paper should be numbered and cross-referenced.
        \item All assumptions should be clearly stated or referenced in the statement of any theorems.
        \item The proofs can either appear in the main paper or the supplemental material, but if they appear in the supplemental material, the authors are encouraged to provide a short proof sketch to provide intuition. 
        \item Inversely, any informal proof provided in the core of the paper should be complemented by formal proofs provided in appendix or supplemental material.
        \item Theorems and Lemmas that the proof relies upon should be properly referenced. 
    \end{itemize}

    \item {\bf Experimental Result Reproducibility}
    \item[] Question: Does the paper fully disclose all the information needed to reproduce the main experimental results of the paper to the extent that it affects the main claims and/or conclusions of the paper (regardless of whether the code and data are provided or not)?
    \item[] Answer: \answerYes{} 
    \item[] Justification: We provided most of the details in order to reproduce the experiments in the Appendix.
    \item[] Guidelines:
    \begin{itemize}
        \item The answer NA means that the paper does not include experiments.
        \item If the paper includes experiments, a No answer to this question will not be perceived well by the reviewers: Making the paper reproducible is important, regardless of whether the code and data are provided or not.
        \item If the contribution is a dataset and/or model, the authors should describe the steps taken to make their results reproducible or verifiable. 
        \item Depending on the contribution, reproducibility can be accomplished in various ways. For example, if the contribution is a novel architecture, describing the architecture fully might suffice, or if the contribution is a specific model and empirical evaluation, it may be necessary to either make it possible for others to replicate the model with the same dataset, or provide access to the model. In general. releasing code and data is often one good way to accomplish this, but reproducibility can also be provided via detailed instructions for how to replicate the results, access to a hosted model (e.g., in the case of a large language model), releasing of a model checkpoint, or other means that are appropriate to the research performed.
        \item While NeurIPS does not require releasing code, the conference does require all submissions to provide some reasonable avenue for reproducibility, which may depend on the nature of the contribution. For example
        \begin{enumerate}
            \item If the contribution is primarily a new algorithm, the paper should make it clear how to reproduce that algorithm.
            \item If the contribution is primarily a new model architecture, the paper should describe the architecture clearly and fully.
            \item If the contribution is a new model (e.g., a large language model), then there should either be a way to access this model for reproducing the results or a way to reproduce the model (e.g., with an open-source dataset or instructions for how to construct the dataset).
            \item We recognize that reproducibility may be tricky in some cases, in which case authors are welcome to describe the particular way they provide for reproducibility. In the case of closed-source models, it may be that access to the model is limited in some way (e.g., to registered users), but it should be possible for other researchers to have some path to reproducing or verifying the results.
        \end{enumerate}
    \end{itemize}

\item {\bf Open access to data and code}
    \item[] Question: Does the paper provide open access to the data and code, with sufficient instructions to faithfully reproduce the main experimental results, as described in supplemental material?
    \item[] Answer: \answerNo 
    \item[] Justification: We planned to release the code upon the acceptance of the paper. Also, we have included most of the details in the Appendix for reproducing the experiments.
    \item[] Guidelines:
    \begin{itemize}
        \item The answer NA means that paper does not include experiments requiring code.
        \item Please see the NeurIPS code and data submission guidelines (\url{https://nips.cc/public/guides/CodeSubmissionPolicy}) for more details.
        \item While we encourage the release of code and data, we understand that this might not be possible, so “No” is an acceptable answer. Papers cannot be rejected simply for not including code, unless this is central to the contribution (e.g., for a new open-source benchmark).
        \item The instructions should contain the exact command and environment needed to run to reproduce the results. See the NeurIPS code and data submission guidelines (\url{https://nips.cc/public/guides/CodeSubmissionPolicy}) for more details.
        \item The authors should provide instructions on data access and preparation, including how to access the raw data, preprocessed data, intermediate data, and generated data, etc.
        \item The authors should provide scripts to reproduce all experimental results for the new proposed method and baselines. If only a subset of experiments are reproducible, they should state which ones are omitted from the script and why.
        \item At submission time, to preserve anonymity, the authors should release anonymized versions (if applicable).
        \item Providing as much information as possible in supplemental material (appended to the paper) is recommended, but including URLs to data and code is permitted.
    \end{itemize}

\item {\bf Experimental Setting/Details}
    \item[] Question: Does the paper specify all the training and test details (e.g., data splits, hyperparameters, how they were chosen, type of optimizer, etc.) necessary to understand the results?
    \item[] Answer: \answerYes{} 
    \item[] Justification: We provided the hyperparameters optimizer etc. in the Appendix.
    \item[] Guidelines:
    \begin{itemize}
        \item The answer NA means that the paper does not include experiments.
        \item The experimental setting should be presented in the core of the paper to a level of detail that is necessary to appreciate the results and make sense of them.
        \item The full details can be provided either with the code, in appendix, or as supplemental material.
    \end{itemize}

\item {\bf Experiment Statistical Significance}
    \item[] Question: Does the paper report error bars suitably and correctly defined or other appropriate information about the statistical significance of the experiments?
    \item[] Answer: \answerYes{} 
    \item[] Justification: we have reported the numerical results under various metrics and provided comparable baselines.
    \item[] Guidelines:
    \begin{itemize}
        \item The answer NA means that the paper does not include experiments.
        \item The authors should answer "Yes" if the results are accompanied by error bars, confidence intervals, or statistical significance tests, at least for the experiments that support the main claims of the paper.
        \item The factors of variability that the error bars are capturing should be clearly stated (for example, train/test split, initialization, random drawing of some parameter, or overall run with given experimental conditions).
        \item The method for calculating the error bars should be explained (closed form formula, call to a library function, bootstrap, etc.)
        \item The assumptions made should be given (e.g., Normally distributed errors).
        \item It should be clear whether the error bar is the standard deviation or the standard error of the mean.
        \item It is OK to report 1-sigma error bars, but one should state it. The authors should preferably report a 2-sigma error bar than state that they have a 96\% CI, if the hypothesis of Normality of errors is not verified.
        \item For asymmetric distributions, the authors should be careful not to show in tables or figures symmetric error bars that would yield results that are out of range (e.g. negative error rates).
        \item If error bars are reported in tables or plots, The authors should explain in the text how they were calculated and reference the corresponding figures or tables in the text.
    \end{itemize}

\item {\bf Experiments Compute Resources}
    \item[] Question: For each experiment, does the paper provide sufficient information on the computer resources (type of compute workers, memory, time of execution) needed to reproduce the experiments?
    \item[] Answer: \answerYes{} 
    \item[] Justification: We provide the hardware setup in the Appendix.
    \item[] Guidelines:
    \begin{itemize}
        \item The answer NA means that the paper does not include experiments.
        \item The paper should indicate the type of compute workers CPU or GPU, internal cluster, or cloud provider, including relevant memory and storage.
        \item The paper should provide the amount of compute required for each of the individual experimental runs as well as estimate the total compute. 
        \item The paper should disclose whether the full research project required more compute than the experiments reported in the paper (e.g., preliminary or failed experiments that didn't make it into the paper). 
    \end{itemize}
    
\item {\bf Code Of Ethics}
    \item[] Question: Does the research conducted in the paper conform, in every respect, with the NeurIPS Code of Ethics \url{https://neurips.cc/public/EthicsGuidelines}?
    \item[] Answer: \answerYes{} 
    \item[] Justification: We are aligned with NeurIPS Ethics Guidelines.
    \item[] Guidelines:
    \begin{itemize}
        \item The answer NA means that the authors have not reviewed the NeurIPS Code of Ethics.
        \item If the authors answer No, they should explain the special circumstances that require a deviation from the Code of Ethics.
        \item The authors should make sure to preserve anonymity (e.g., if there is a special consideration due to laws or regulations in their jurisdiction).
    \end{itemize}

\item {\bf Broader Impacts}
    \item[] Question: Does the paper discuss both potential positive societal impacts and negative societal impacts of the work performed?
    \item[] Answer: \answerNo{} 
    \item[] Justification: This is a proof of concept paper. Broader Impacts is not applicable.
    \item[] Guidelines:
    \begin{itemize}
        \item The answer NA means that there is no societal impact of the work performed.
        \item If the authors answer NA or No, they should explain why their work has no societal impact or why the paper does not address societal impact.
        \item Examples of negative societal impacts include potential malicious or unintended uses (e.g., disinformation, generating fake profiles, surveillance), fairness considerations (e.g., deployment of technologies that could make decisions that unfairly impact specific groups), privacy considerations, and security considerations.
        \item The conference expects that many papers will be foundational research and not tied to particular applications, let alone deployments. However, if there is a direct path to any negative applications, the authors should point it out. For example, it is legitimate to point out that an improvement in the quality of generative models could be used to generate deepfakes for disinformation. On the other hand, it is not needed to point out that a generic algorithm for optimizing neural networks could enable people to train models that generate Deepfakes faster.
        \item The authors should consider possible harms that could arise when the technology is being used as intended and functioning correctly, harms that could arise when the technology is being used as intended but gives incorrect results, and harms following from (intentional or unintentional) misuse of the technology.
        \item If there are negative societal impacts, the authors could also discuss possible mitigation strategies (e.g., gated release of models, providing defenses in addition to attacks, mechanisms for monitoring misuse, mechanisms to monitor how a system learns from feedback over time, improving the efficiency and accessibility of ML).
    \end{itemize}
    
\item {\bf Safeguards}
    \item[] Question: Does the paper describe safeguards that have been put in place for responsible release of data or models that have a high risk for misuse (e.g., pretrained language models, image generators, or scraped datasets)?
    \item[] Answer: \answerNA{} 
    \item[] Justification: This is a proof of concept paper. The safeguards is not applicable.
    \item[] Guidelines:
    \begin{itemize}
        \item The answer NA means that the paper poses no such risks.
        \item Released models that have a high risk for misuse or dual-use should be released with necessary safeguards to allow for controlled use of the model, for example by requiring that users adhere to usage guidelines or restrictions to access the model or implementing safety filters. 
        \item Datasets that have been scraped from the Internet could pose safety risks. The authors should describe how they avoided releasing unsafe images.
        \item We recognize that providing effective safeguards is challenging, and many papers do not require this, but we encourage authors to take this into account and make a best faith effort.
    \end{itemize}

\item {\bf Licenses for existing assets}
    \item[] Question: Are the creators or original owners of assets (e.g., code, data, models), used in the paper, properly credited and are the license and terms of use explicitly mentioned and properly respected?
    \item[] Answer: \answerYes{} 
    \item[] Justification: The code is original. For the data, we disclosed the original owners.
    \item[] Guidelines:
    \begin{itemize}
        \item The answer NA means that the paper does not use existing assets.
        \item The authors should cite the original paper that produced the code package or dataset.
        \item The authors should state which version of the asset is used and, if possible, include a URL.
        \item The name of the license (e.g., CC-BY 4.0) should be included for each asset.
        \item For scraped data from a particular source (e.g., website), the copyright and terms of service of that source should be provided.
        \item If assets are released, the license, copyright information, and terms of use in the package should be provided. For popular datasets, \url{paperswithcode.com/datasets} has curated licenses for some datasets. Their licensing guide can help determine the license of a dataset.
        \item For existing datasets that are re-packaged, both the original license and the license of the derived asset (if it has changed) should be provided.
        \item If this information is not available online, the authors are encouraged to reach out to the asset's creators.
    \end{itemize}

\item {\bf New Assets}
    \item[] Question: Are new assets introduced in the paper well documented and is the documentation provided alongside the assets?
    \item[] Answer: \answerNA{} 
    \item[] Justification: It is not applicable for our paper.
    \item[] Guidelines:
    \begin{itemize}
        \item The answer NA means that the paper does not release new assets.
        \item Researchers should communicate the details of the dataset/code/model as part of their submissions via structured templates. This includes details about training, license, limitations, etc. 
        \item The paper should discuss whether and how consent was obtained from people whose asset is used.
        \item At submission time, remember to anonymize your assets (if applicable). You can either create an anonymized URL or include an anonymized zip file.
    \end{itemize}

\item {\bf Crowdsourcing and Research with Human Subjects}
    \item[] Question: For crowdsourcing experiments and research with human subjects, does the paper include the full text of instructions given to participants and screenshots, if applicable, as well as details about compensation (if any)? 
    \item[] Answer: \answerNA{} 
    \item[] Justification: This is no human subjects included in this paper.
    \item[] Guidelines:
    \begin{itemize}
        \item The answer NA means that the paper does not involve crowdsourcing nor research with human subjects.
        \item Including this information in the supplemental material is fine, but if the main contribution of the paper involves human subjects, then as much detail as possible should be included in the main paper. 
        \item According to the NeurIPS Code of Ethics, workers involved in data collection, curation, or other labor should be paid at least the minimum wage in the country of the data collector. 
    \end{itemize}

\item {\bf Institutional Review Board (IRB) Approvals or Equivalent for Research with Human Subjects}
    \item[] Question: Does the paper describe potential risks incurred by study participants, whether such risks were disclosed to the subjects, and whether Institutional Review Board (IRB) approvals (or an equivalent approval/review based on the requirements of your country or institution) were obtained?
    \item[] Answer: \answerNA{} 
    \item[] Justification: This is not applicable for this paper.
    \item[] Guidelines:
    \begin{itemize}
        \item The answer NA means that the paper does not involve crowdsourcing nor research with human subjects.
        \item Depending on the country in which research is conducted, IRB approval (or equivalent) may be required for any human subjects research. If you obtained IRB approval, you should clearly state this in the paper. 
        \item We recognize that the procedures for this may vary significantly between institutions and locations, and we expect authors to adhere to the NeurIPS Code of Ethics and the guidelines for their institution. 
        \item For initial submissions, do not include any information that would break anonymity (if applicable), such as the institution conducting the review.
    \end{itemize}

\end{enumerate}

\end{document}